\documentclass[epj]{svjour}
\onecolumn \normalsize
\newcommand{\pt}{$p_T$}
\usepackage{graphics}
\begin{document}
\title{Identified hadron production at high transverse momenta in p+p collisions at $\sqrt{s_{NN}}$ = 200
GeV in STAR} %\subtitle{Do you have a subtitle?\\ If so, write it here}
\author{Yichun Xu\inst{1} (for the STAR collaboration) %\and Second author\inst{2}% etc
\thanks{\emph{Present address:} xuyichun@rcf.rhic.bnl.gov}%
}                     % Do not remove
\institute{Department of Modern Physics, University of Science and
Technology of China, Hefei, Anhui 230026, China \and Physics
Department, Brookhaven National Laboratory, Upton, NY 11973, USA}
\date{Received: date / Revised version: date}
\abstract{ We report the transverse momentum ($p_{T}$)
distributions for identified charged pions, protons and
anti-protons using events triggered by high deposit energy in the
Barrel Electro-Magnetic Calorimeter (BEMC) from $p$ + $p$
collisions at $\sqrt{s_{NN}}$ = 200 GeV. The spectra are measured
around mid-rapidity ($\mid$ y $\mid$ $<$ 0.5) over the range of 3
$< p_{T} <$ 15 GeV/$c$ with particle identification (PID) by the
relativistic ionization energy loss ($rdE/dx$) in the Time
Projection Chamber (TPC) in the Solenoidal Tracker at RHIC (STAR).
The charged pion, proton and anti-proton spectra at high $p_{T}$
are compared with published results from minimum bias triggered
events and the Next-Leading-Order perturbative quantum
chromodynamic (NLO pQCD) calculations (DSS, KKP and AKK 2008). In
addition, we present the particle ratios of $\pi^{-}/\pi^{+}$,
$\overline{p}/p$, $p/\pi^{+}$ and $\overline{p}/\pi^{-}$ in $p+p$
collisions.
 % end of PACS codes
} %end of abstract
%
%\title{PID spectra}
%\author{Yichun Xu (for the STAR Collaboration)}
\maketitle
\section{Introduction}
\label{intro} The study of identified hadron ($\pi^{\pm}$,
$K^{\pm}$, $p(\overline{p}$)) spectra at high $p_T$ in p+p
collisions provides a good test of perturbative quantum
chromodynamics (pQCD) \cite{pQCD}. In different NLO pQCD
calculations, the inclusive production of single hadron is
described by the convolution of parton distribution functions
(PDFs), parton interaction cross-sections, and fragmentation
functions (FFs) which are parameterized by measured hadron
spectra. From the minimum-bias triggered $p+p$ collisions in the
year 2003, the $p(\overline{p}$) and charged pion spectra were
measured at $p_{T} \leq$ 7 GeV/$c$ and $p_{T} \leq$ 10 GeV/$c$
with significant systematic errors due to the uncertainties in
mean $dE/dx$ position for protons and kaons \cite{starppPID}. In
order to understand mechanism of hadron production, it is
necessary to make a strict constraint on the quark and gluon FFs
by comparing theory with experimental data. In addition, it's also
a good baseline for studying color charge effect of parton energy
loss in heavy ion collisions, in which hadron spectra were
measured up to 12 GeV/$c$ \cite{starAuAuPID}.

In this article, we present the $p_{T}$ spectra for identified
charge pions, protons and anti-protons in p + p collisions at
$\sqrt{s_{NN}}$ = 200 GeV by the STAR experiment at RHIC. The
results will be compared with NLO pQCD calculations. The particle
ratios of $\pi^{-}/\pi^{+}$, $\overline{p}/p$, $p/\pi^{+}$ and
$\overline{p}/\pi^{-}$ in $p+p$ collisions will be presented and
and compared with results in $d+Au$ collisions
\cite{starppPID,stardAuPID}.
\section{Experiment and Analysis}
The STAR main tracking detector, the TPC, covering full azimuthal
angle (2$\pi$) and $\mid \eta \mid$ $<$ 1.3 in pseudo-rapidity
provides a way to identify charged hadrons by measuring momentum
and $dE/dx$ information of charged particles. In addition, the
BEMC covering 2$\pi$ azimuthal angle and 0 $<$ $\eta$ $<$ 1 in
year 2005 provides deposited energy of electron, and
electro-magnetic shower shape, size and position, which are used
to enhance electron yields relative to other hadrons. This is
helpful for re-calibrating $rdE/dx$ \cite{reCalibrationMethod},
which will be discussed later.

Due to empirical parameters in theory, gas multiplication gains
and noise of the TPC electronics, and pileup in high luminosity
environment, measured $rdE/dx$ values are deviated from
theoretical predictions. Dominant pion yields shadow kaons and
protons in the $rdE/dx$ distribution, although the $rdE/dx$
separations among $\pi^{\pm}$, $K^{\pm}$, and p($\overline{p}$)
are about 1-3$\sigma$. This results in large systematic errors due
to the uncertainty of $rdE/dx$ positions. Knowledge of the precise
$rdE/dx$ position for those hadrons is important to understand the
efficiencies of particle identification (PID) selection and to
reduce the systematic uncertainty in identified hadron yields. In
order to improve the PID at high \pt, the re-calibration method
\cite{reCalibrationMethod} is used to locate the $rdE/dx$
positions for different charged particles with good precision. To
separate pion and electron clearly and get $rdE/dx$ of electron
precisely, information of the BEMC is used to get electron
enhancement data sample. Meanwhile, Strange hadrons
$\Lambda\rightarrow p+\pi^{-}$ ($\overline{\Lambda}\rightarrow
\overline{p}+\pi^{+}$) and $K^{0}_{S}\rightarrow\pi^{+}+\pi^{-}$
are reconstructed by their decay topology to identify their decay
daughters in the TPC. This provides $rdE/dx$ peak position from
samples of pure protons and charged pions
\cite{reCalibrationMethod}.

\begin{figure}
\resizebox{1.0\textwidth}{!}{
\includegraphics{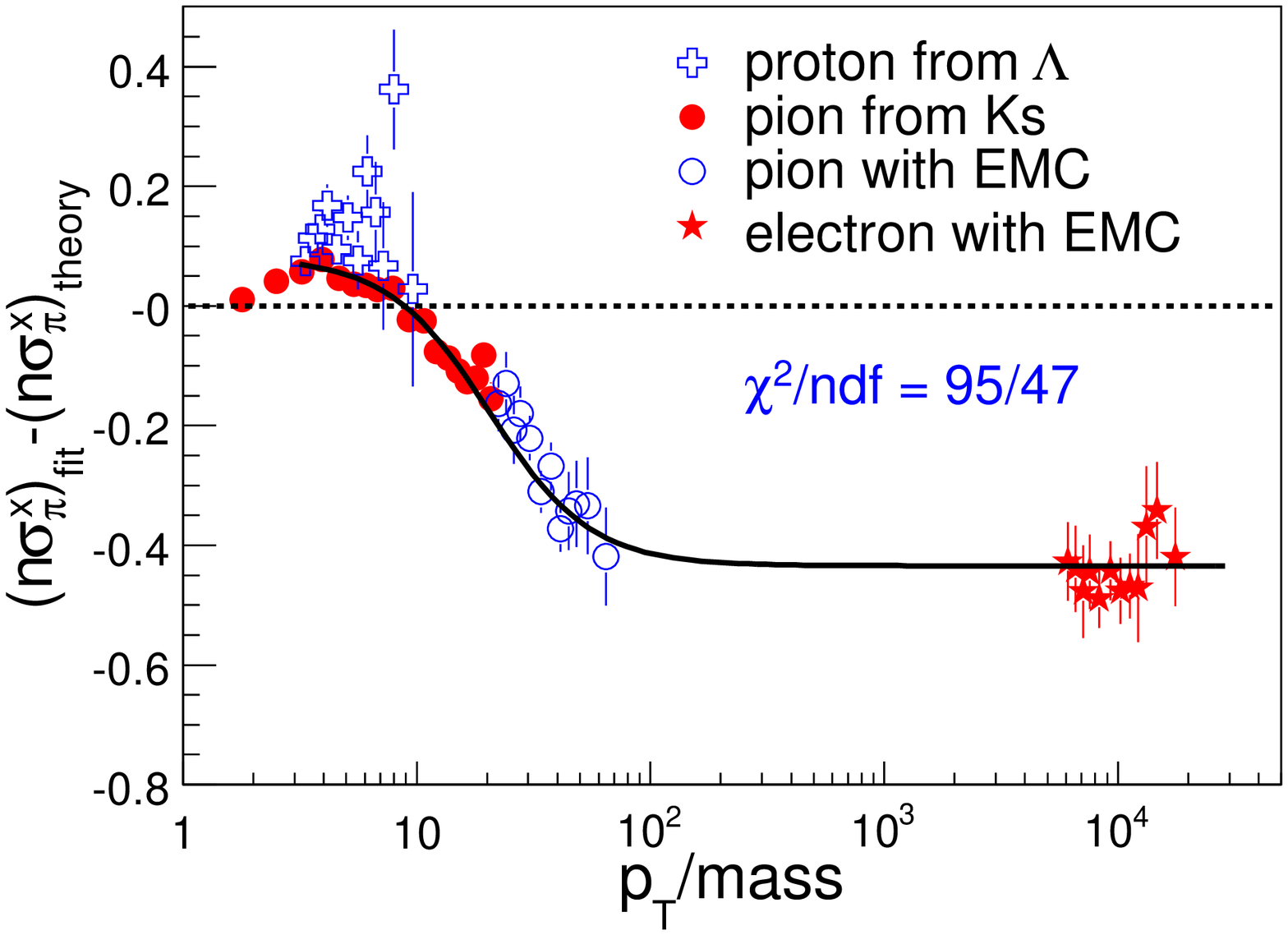}
\includegraphics{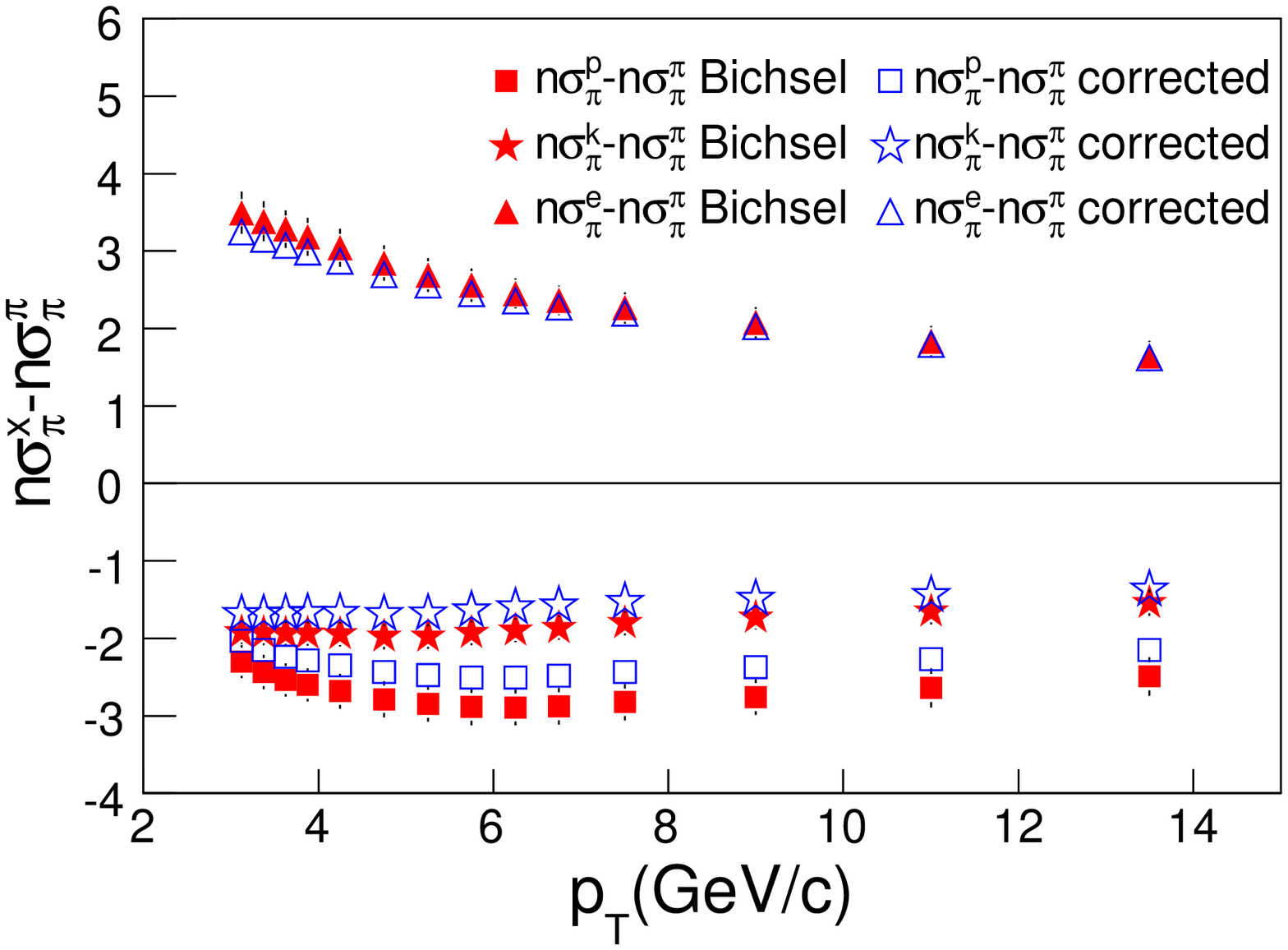}
} \caption{On the left panel, the $dE/dx$ deviation in $\sigma$ as
functions of transverse momentum divided by mass (\pt/mass). The
blue crosses are proton decayed from $\Lambda$, the red filled
dots are pion from $K^{0}_{S}$, and blue open circles and red
stars are pion and electron from electron enhancement sample
respectively. On the right panel, comparison of the relative
$dE/dx$ peak position of $n\sigma_{\pi}^{K}$, $n\sigma_{\pi}^{p}$,
$n\sigma_{\pi}^{e}$. All solid dots depict theoretical values, and
open ones are re-calibrated results.}
\label{recalibration}       % Give a unique label
\end{figure}

With these precise $rdE/dx$ information, the left panel of
Figure~\ref{recalibration} shows the deviation of $rdE/dx$
normalized by pion $rdE/dx$ between data and theoretical values.
This distribution is described by a function, $f(x) =
A+\frac{B}{C+x^{2}}$. Meanwhile, the separation between identified
hadrons and pions are derived from this fit function, as shown on
the right panel of Figure~\ref{recalibration}.

\begin{figure}
\resizebox{0.9\textwidth}{!}{
\includegraphics{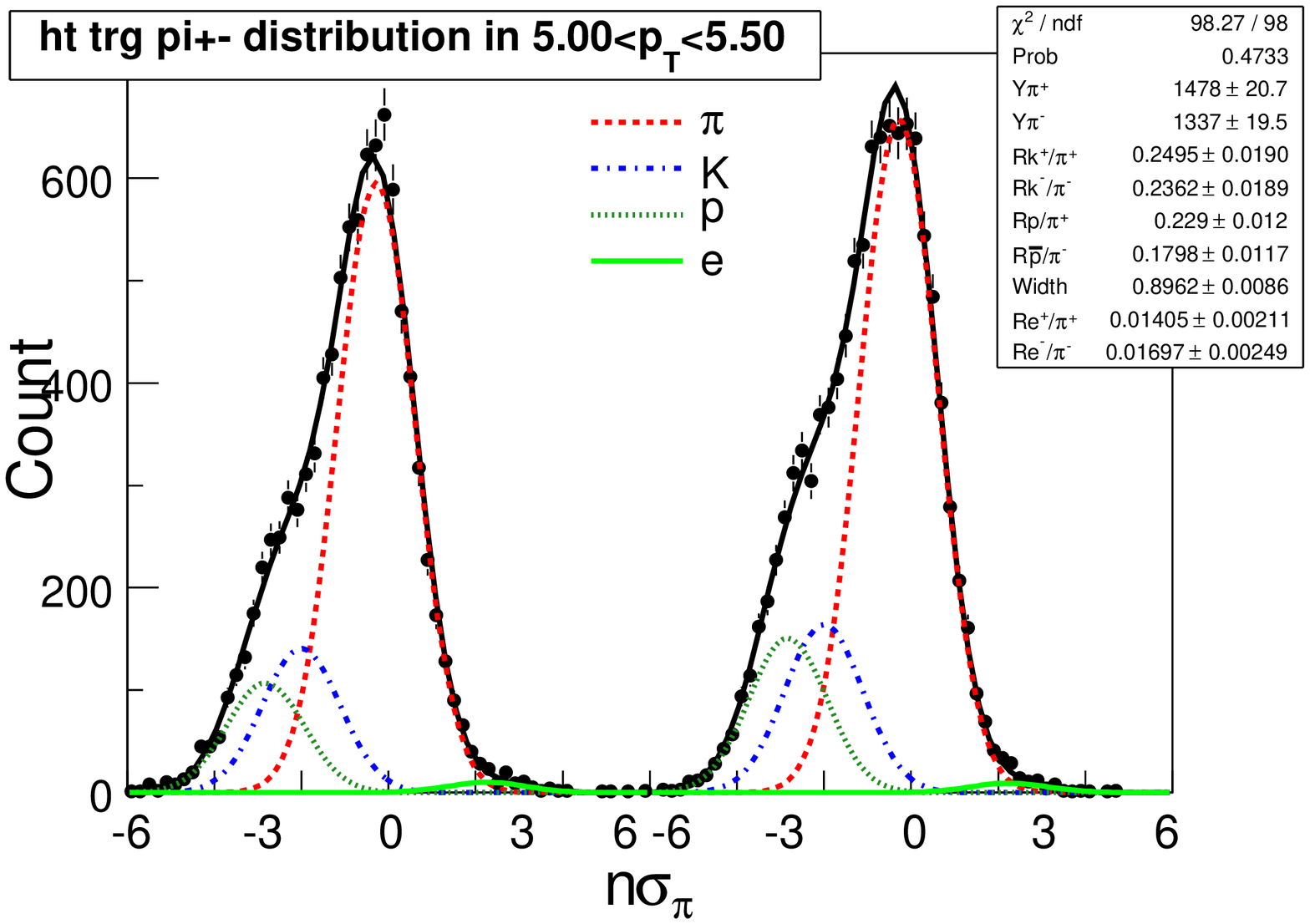}
\includegraphics{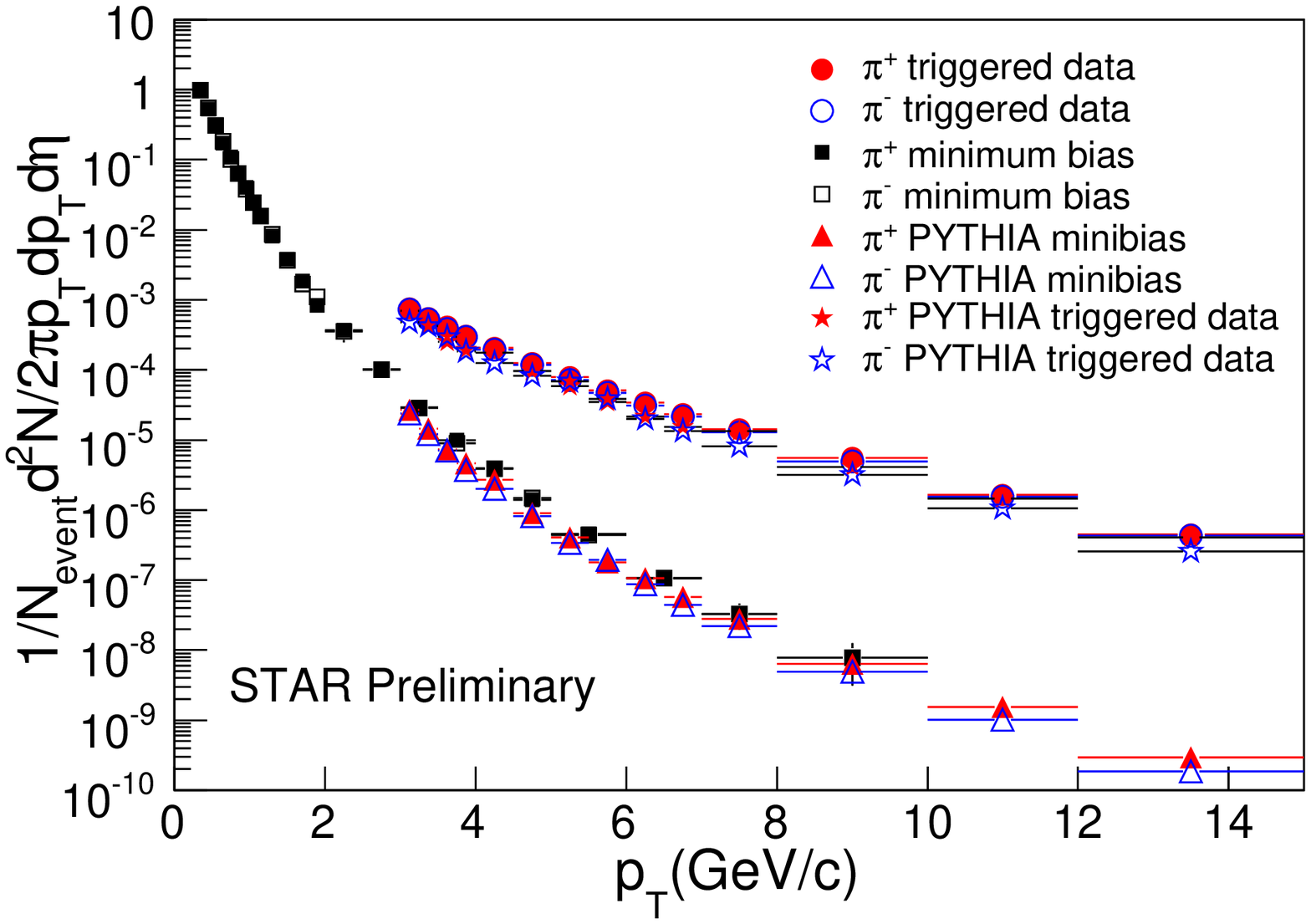}
} \caption{On the left panel, $n\sigma_{\pi}^{h}$ distribution at
3.75 $< p_{T} <$ 4.0 GeV/$c$ for positive (right part) and
negative (left part) particles. The black line is the fit curve by
8-Gaussian function, including pion (red dashed line), kaon (blue
dot-dashed line), proton (green dotted line), and electron (green
solid line). On the right panel, pion spectra from minimum bias
events, BEMC triggered events, and simulated minimum bias and BEMC
triggered events.}
\label{pionspectra}       % Give a unique label
\end{figure}

In order to extract charged pion and proton (anti-proton) yields,
a total of 5.6 million BEMC triggered events in year 2005 (with
transverse energy $E_{T}$ $>$ 6.4 GeV) have been analyzed. The
details of BEMC trigger condition can be found in Ref.
\cite{BEMC}. Since pions are the dominant component of the
hadrons, pion raw yields can be derived by fitting $rdE/dx$
distribution using 8-Gaussian with four fixed re-calibrated
parameters for peak positions. The left panel of
Figure~\ref{pionspectra} shows the $rdE/dx$ distribution at 5.0
$<$ $p_{T}$ $<$ 5.5 GeV/c and $\mid \eta \mid$ $<$ 0.5. For proton
yields, we used two methods. One method is based on track-by-track
selection using a cut in $rdE/dx$. The other method is fitting
$dE/dx$ distribution with fixed proton peak position. The final
yields we used in this analysis are averaged from these two
methods. Half of the difference ($<$ 13$\%$) between them is
included in systematic uncertainty.

The right panel on Figure~\ref{pionspectra} shows that there is
great difference between invariant yields of charged pions from
BEMC triggered events (red and blue circles) and minimum bias
events (black squares) because of trigger enhancement. In order to
correct this effect, PYTHIA events are embedded in GEANT with STAR
geometry, which can simulate the realistic response of the STAR
detector in experiment, including signal of read-out and response
of electronics, when tracks are propagated through detectors. With
simulated signal, different triggered events are selected by
passing different detector thresholds as real events in STAR
experiment. On the right panel in Figure \ref{pionspectra}, stars
and triangles represent charged pion spectra from simulated BEMC
triggered and minimum bias events.

\begin{figure}
\resizebox{0.9\textwidth}{!}{
\includegraphics{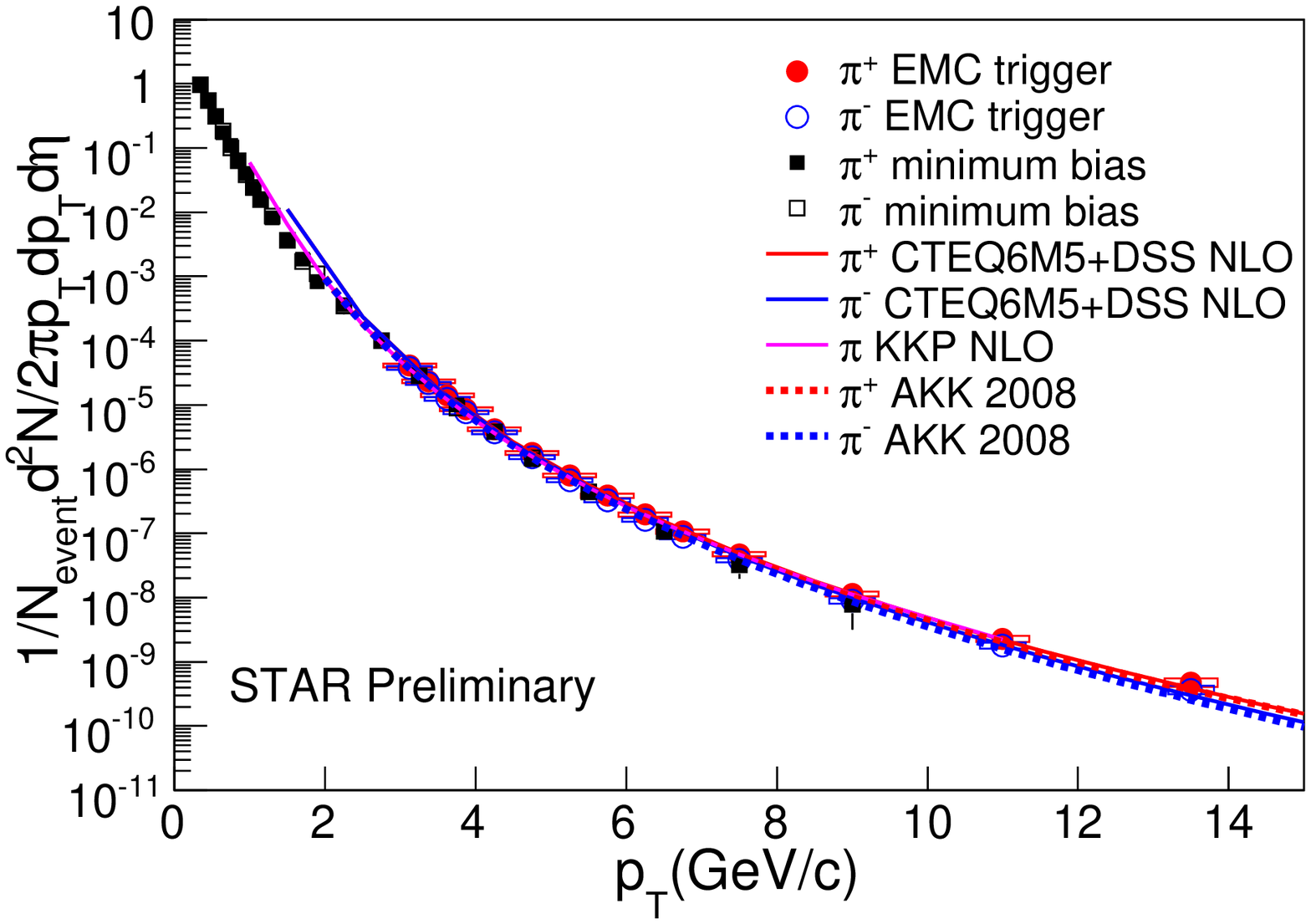}
\includegraphics{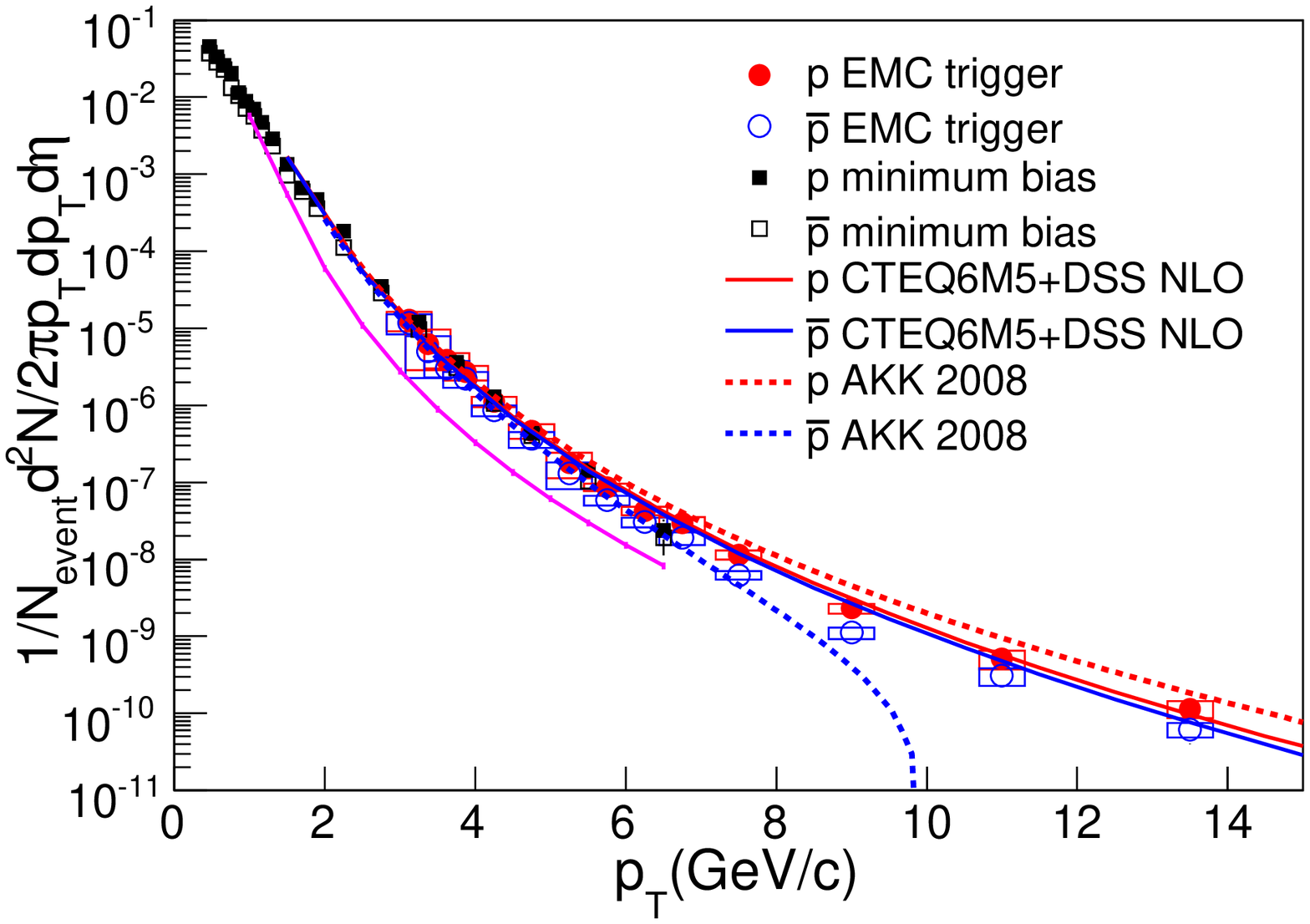}
} \caption{Spectra for pion (on the left panel) and proton (on the
right panel) with NLO pQCD calculations including DSS, AKK 2008.}
\label{pionprotonspectra}       % Give a unique label
\end{figure}
With the simulation, we can calculate the enhancement factors
versus $p_{T}$ by dividing BEMC triggered spectra by minimum bias
triggered spectra. With the trigger enhancement factors and
tracking efficiencies ($\sim$88$\%$ at 3 $<$ $p_{T}$ $<$ 15
GeV/$c$), the charged pion spectra are corrected and shown on the
left panel of Figure \ref{pionprotonspectra}. The corrected
spectra of pions are consistent with minimum bias results at the
overlapped $p_{T}$ range, and the NLO pQCD calculations. With
$p/\pi^{+}$ and $\overline{p}/\pi^{-}$ ratios, the invariant
yields of proton (anti-proton) are calculated by pion spectra
times $p/\pi^{+}$ ($\overline{p}/\pi^{-}$)ratios. The spectra of
protons and anti-protons are shown on the right panel in Figure
\ref{pionprotonspectra}. Almost all these NLO pQCD predictions can
not describe our proton and anti-proton spectra well at high \pt,
especially AKK 2008 for anti-proton, which is going down to zero
at 9 GeV/$c$ (from S. Albino private communication). The AKK 2008
\cite{AKK2008} includes BRAHMS data points at high rapidity (2.95
$<$ y $<$ 3.1) \cite{BRAHMS} for parametrization. Our data will
provide a good constraint for the NLO pQCD calculation.

The total systematic uncertainties associated with pion are
estimated to be less than 15$\%$. The systematic uncertainty
consists of uncertainty of peak position ($<$4$\%$), charge
distortion ($<$12$\%$), momentum resolution ($<$5$\%$) and
efficiencies ($<$5$\%$). Proton spectra have similar systematic
uncertainty sources. In addition, two methods to getting proton
yields results
in $\sim$13$\%$ contribution. %Table~\ref{tableSyst} shows all the
%sources of systematic uncertainties for charged pions, protons and
%anti-protons.

%\begin{table}
%\begin{center}
%\caption{uncertainties contributions for pion and proton}
%\label{tableSyst}       % Give a unique label
%\begin{tabular}{llllll}
%\hline\noalign{\smallskip}
%species & peak position & efficiency & momentum resolution & charge distortion  & method\\
%pion & $<$4$\%$ & 5$\%$ & $<$5$\%$ & $<$12$\%$ & -\\
%p($\overline{p}$) & $<$16$\%$ & 5$\%$ & $<$5$\%$ & $<12\%$  &$<20\%$\\
%\end{tabular}
%\end{center}
%\end{table}

\begin{figure}
\resizebox{0.9\textwidth}{!}{
\includegraphics{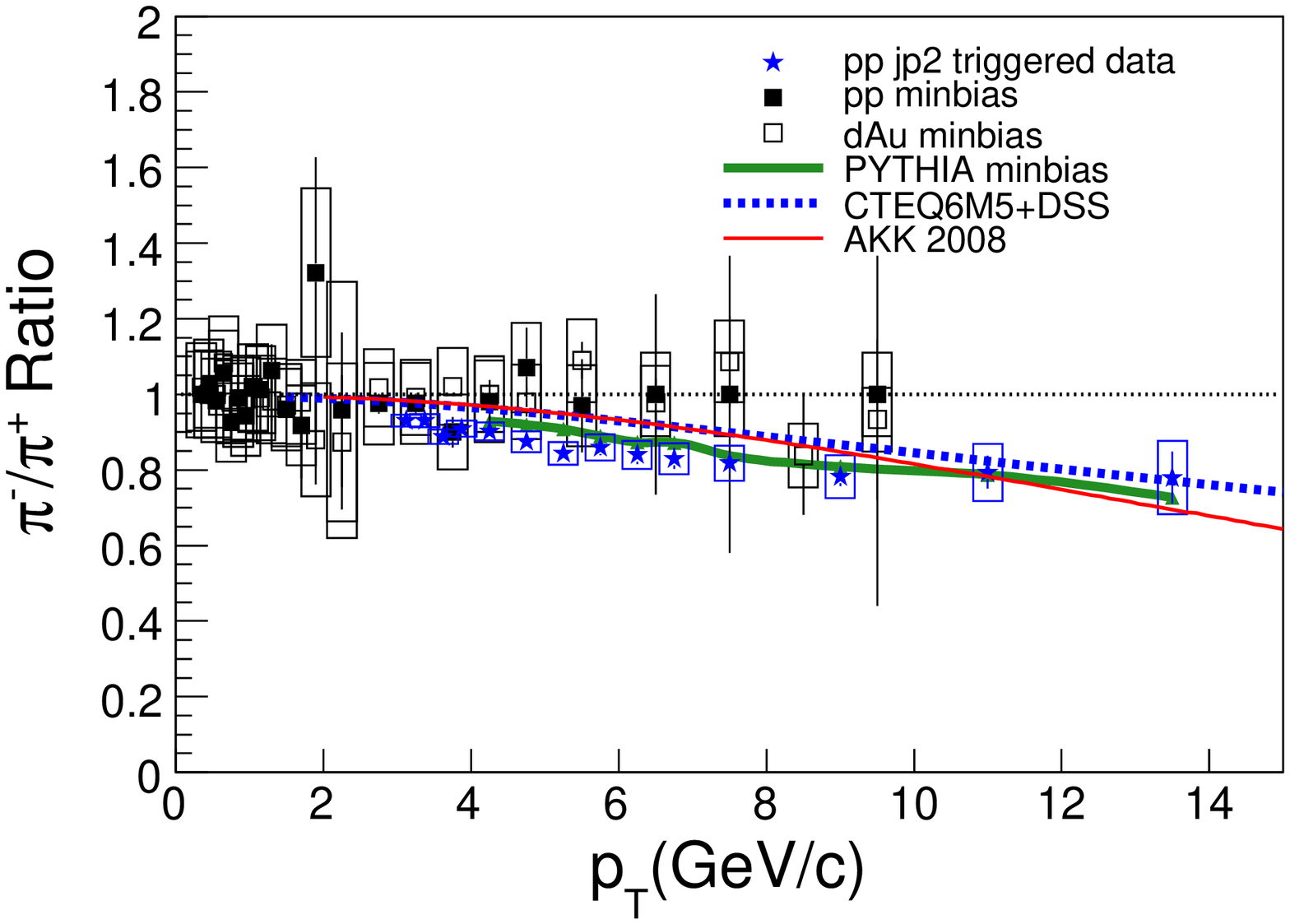}
\includegraphics{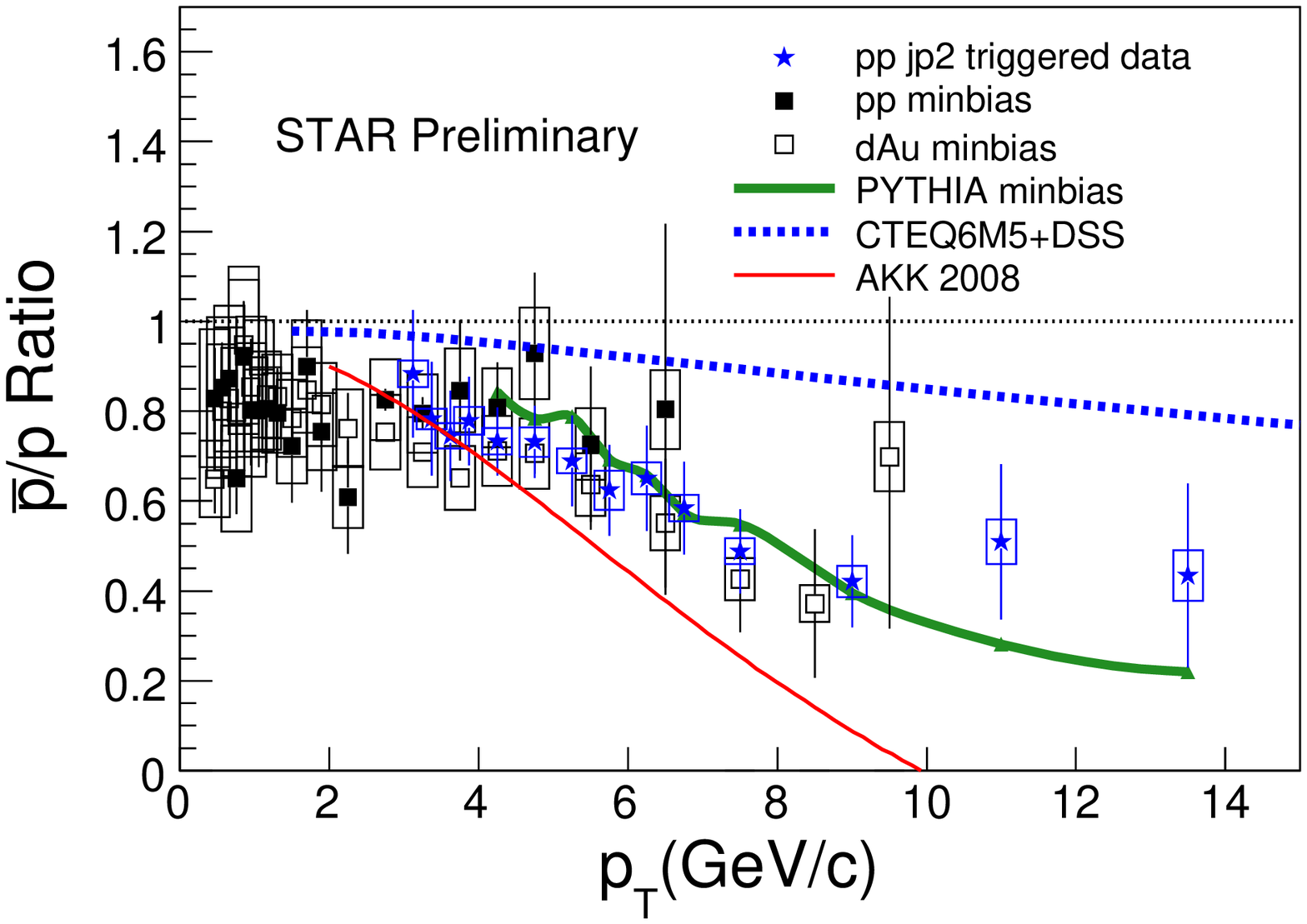}
}
%\end{figure}
%\begin{figure}
\resizebox{0.9\textwidth}{!}{
\includegraphics{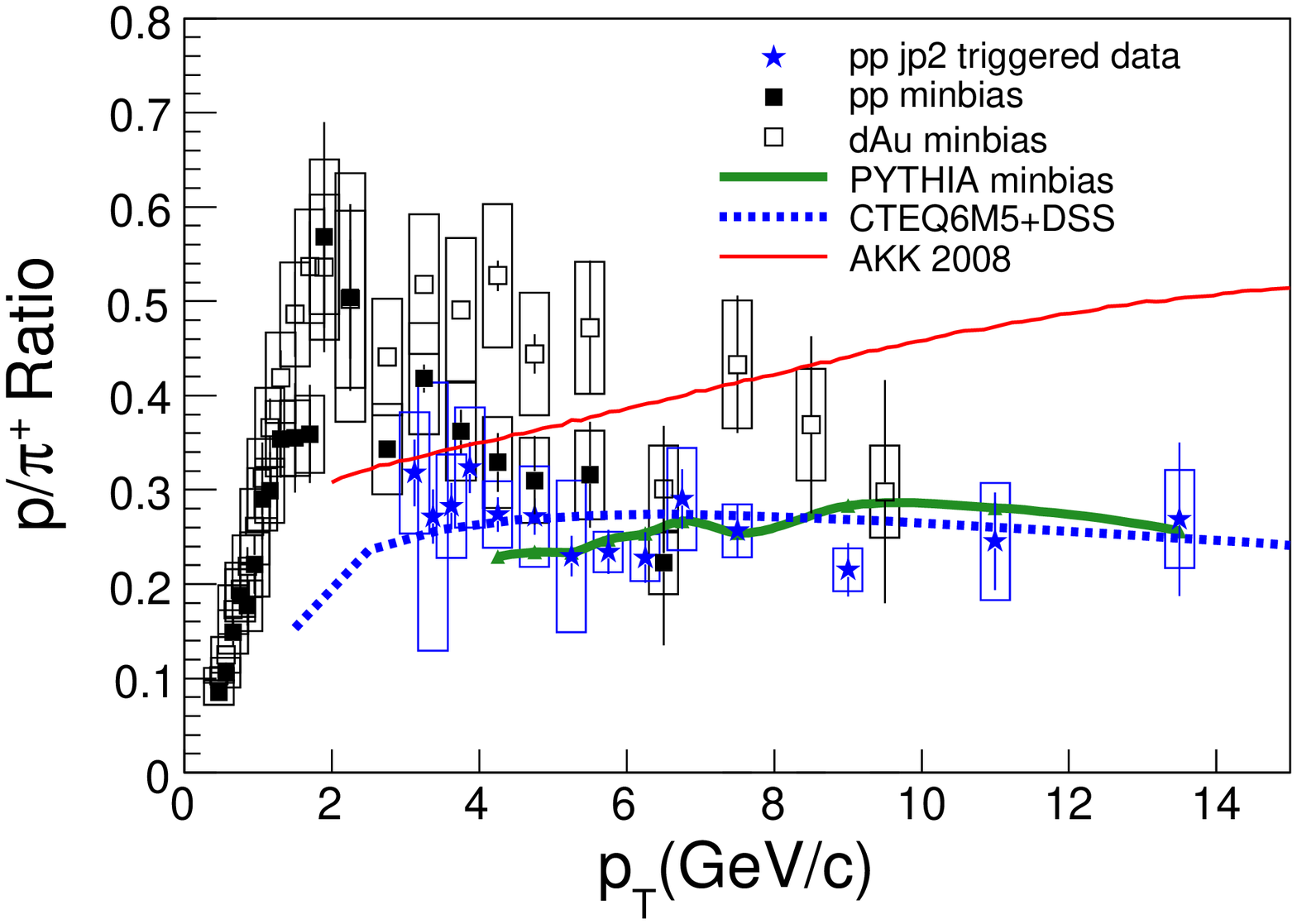}
\includegraphics{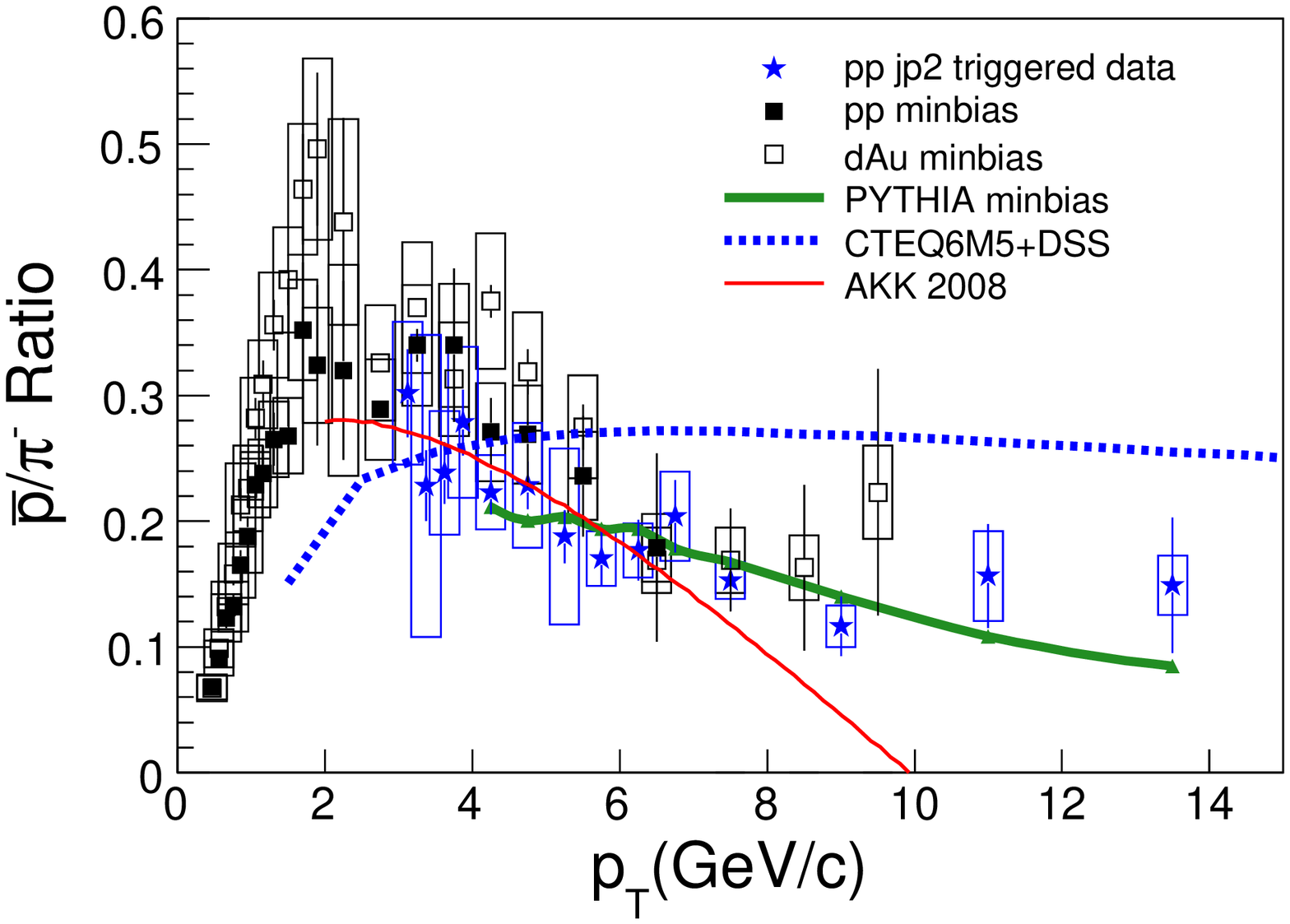}
}\caption{Ratios of $\pi^{-}$/$\pi^{+}$, $\overline{p}$/p,
p/$\pi^{+}$ and $\overline{p}$/$\pi^{-}$ from minimum bias events
(black squares), BEMC triggered events (blue stars), and PYTHIA
simulation (green solid line), DSS (blue dashed-line) and AKK 2008
(red solid line) predictions.}
\label{ratios}       % Give a unique label
\end{figure}

The particle ratios at mid-rapidity as a function of $p_{T}$ are
shown on Figure~\ref{ratios}, and compared with published results
from minimum bias $p+p$ and $d+Au$ collisions. The predictions
from different model, such as PYTHIA, DSS, and AKK 2008 are also
shown. All the boxes and bars represent systematical and
statistical errors. Our results are consistent with those from
minimum bias $p+p$ collisions. For the models, only PYTHIA can
describe these ratios well, DSS over-predicts anti-proton, AKK
2008 over-predicts proton, and under-predicts anti-proton. The
$\pi^{-}/\pi^{+}$ ratio from BEMC triggered events decreases with
increasing \pt, which indicates significant valence quark
contribution to charge pion production. The decrease of
$\overline{p}$/p ratio also indicates a significant quark
contribution to baryon production. The $p/\pi^{+}$ and
$\overline{p}/\pi^{-}$ ratios decrease at intermediate $p_{T}$
range (2 $<$ $p_{T}$ $<$6 GeV/$c$) and approach constant at high
\pt, which are consistent with PYTHIA simulation. The $p/\pi^{+}$
ratio in $p+p$ collision is lower than that in $d+Au$ collisions
in the intermediate $p_{T}$ range.

%\label{sec:1}
%and \cite{RefJ}
%\subsection{Method}
%\label{sec:2}
%as required. Don't forget to give each section
%and subsection a unique label (see Sect.~\ref{sec:1}).
%\subsection{Results}
\section{Summary and discussion}
We have presented transverse momentum spectra for identified
charged pions, protons and anti-protons from $p+p$ collisions
triggered by the BEMC at $\sqrt{s_{NN}}$ = 200 GeV. With PYTHIA
and GEANT simulation, the transverse momentum spectra around
mid-rapidity ($\mid \eta \mid$ $<$ 0.5) are extended up to $p_{T}$
$>$ 15 GeV/$c$ with PID by the $rdE/dx$ in the TPC. Comparison of
spectra to NLO pQCD predictions, DSS and AKK 2008 can describe
pion spectra well, but poorly for proton and anti-proton spectra,
especially AKK 2008 for anti-proton. Our data can provide a good
constraint to the pQCD calculations, and furthermore, understand
quark and gluon contributions. Ratios of $\pi^{-}$/$\pi^{+}$,
$\overline{p}/p$, $p/\pi^{+}$ and $\overline{p}$/$\pi^{-}$ are
consistent with PYTHIA, but inconsistent with NLO pQCD
calculations, i.e. DSS over-predict anti-protons, AKK 2008
over-predict proton, and under-predict anti-proton. In addition,
the decrease of $\pi^{-}$/$\pi^{+}$ and $\overline{p}/p$ indicate
significant quark contribution to hadrons.

% BibTeX users please use
% \bibliographystyle{}
% \bibliography{}

%
%% For one-column wide figures use
%\begin{figure}
%% Use the relevant command for your figure-insertion program
%% to insert the figure file.
%% For example, with the option graphics use
%\resizebox{0.75\textwidth}{!}{%
%  \includegraphics{leer.eps}
%}
%% If not, use
%%\vspace{5cm}       % Give the correct figure height in cm
%\end{figure}
%%
%% For two-column wide figures use
%\begin{figure*}
%% Use the relevant command for your figure-insertion program
%% to insert the figure file. See example above.
%% If not, use
%\vspace*{5cm}       % Give the correct figure height in cm
%\end{figure*}
%\begin{table}
%\begin{tabular}{lll}
%\hline\noalign{\smallskip}
%\end{tabular}
%% Or use
%\vspace*{5cm}  % with the correct table height
%\end{table}
%

\end{document}